\documentstyle[12pt,epsf]{article}
\newcommand{\vect}[1]{\mbox{\bf #1}}

\newcommand{\rms}[1]{\mbox{\scriptsize #1}}

\newcommand{\lw}[1]{\smash{\lower 1.5ex\hbox{#1}}}

\begin{document}

\title{\bf 
Analysis of Bose-Einstein correlations \\
in $e^+e^- \to W^+W^-$ events \\
including final state interactions}
\author{M. Biyajima, S. Sano and T. Osada${^1}$\thanks{
e-mail: osada@fma1.if.usp.br}\\
Department of Physics, Shinshu University, Matsumoto-390, Japan\\
${^1}$Instituto de Fisica, Universidade de S\~ao Paulo-SP, Brazil
}
\date{\today}
\maketitle

\baselineskip=5.5mm
\begin{abstract}
Recently DELPHI Collaboration reported new data on Bose-Einstein
correlations (BEC) measured in ${e^+e^- \to W^+W^-}$ events.
Apparently no enhancement has been observed. We have analyzed these
data including final state interactions (FSI) of both Coulomb and
strong ($s$-wave) origin and found that there is enhancement in BEC
but it is overshadowed by the FSI which are extremely important for
those events. We have found the following values for the size of the
interaction range $\beta$ and the degree of coherence $\lambda$:
$\beta$=0.87 $\pm 0.31$~fm and  $\lambda$=1.19~$\pm 0.48$, respectively.
\end{abstract}

\noindent 
{\bf 1.~Introduction:~~} Recently DELPHI Collaboration has reported 
an interesting result on the Bose-Einstein correlations (BEC) in 
${e^+e^- \to W^+W^-}$ events at $ \surd{s} = 172 $GeV \cite{del97}.
Namely, analyzing their data by the following formula
\begin{eqnarray}
     N^{\pm\pm}/N^{BG(+-)} = c~[~1 + \lambda \exp(-\beta^2Q^2/2)~],
     \label{standard-formula}
\end{eqnarray}
where the denominator $N^{BG(+-)}$ denotes the contribution of the 
different charge pair, 
($c$ and $\lambda$ are the normalization factor and the degree
of coherence, respectively), 
authors of Ref.~\cite{del97} found that\\
$${   c = 1.0~~,  ~~\lambda = -0.22 \pm 0.2 \mbox{(statics)}, 
~~\mbox{and}  ~~\beta = 0.5~\mbox{fm},} $$ 
i.e., apparently there is no enhancement due to the Bose-Einstein
effect in $e^+e^- \to W^+W^-$ events, most probably because measured
pions come from different $W's$ (i.e., different "sources").\\ 

On the other hand, different approach(weight function method) 
to the BEC in $e^+e^- \to W^+W^-$ events used recently 
in Ref.\cite{kartvel} lead to slightly
different results, namely  $\lambda$ was estimated to be equal $0.032
\sim 0.146$\footnote{Another scheme (shift of the final state momentum) 
is recently argued in Ref.\cite{Lonnbland-1998} in 
order to also determine the $W$-mass.} (See Table I).\\   

Because in these reactions the space-time distances between $W$'s are
very small (of the order $0.5$ fm) the usual Coulombic and strong
final state interactions(FSI) between $\pi^{\pm}$-$\pi^{\pm}$ pairs
should be specially important here and should be taken into account
in order to estimate correctly and precisely $\beta$ and  $\lambda$.
Therefore, we would like to re-analyze data of \cite{del97} by
formula containing the FSI (of both the Coulombic and strong origin,
the later in $s$-wave) proposed by us in Ref.~\cite{osb96}.\\

In the next section, a theoretical formula for correlation of
$\pi^{\pm}$-$\pi^{\pm}$ pairs are derived. In Sec. 3 we derive also
a theoretical formula for $\pi^{+}$-$\pi^{-}$ pairs. Using these two
formulas we analyze in Sec. 4 $e^+e^- \to W^+W^-$ events showing
clearly the roles of strong and Coulomb final state interactions 
in the BEC. We also predict in this Section correlation functions for
$N^{\pm\pm}/N^{BG}$ and $N^{+-}/N^{BG}$. Our concluding remarks are
presented in the last Section.\\

~\\
\noindent 
{\bf 2.~~A theoretical formula for $\pi^{\pm}\pi^{\pm}$ pairs 
including the FSI:}\\
To describe correlations of pair of identical bosons we need to
symmetrize the corresponding total production amplitude, which reads
now: 
\begin{equation}
  A_{12} = \frac{1}{\sqrt{2}}
 [ \Psi_{\rms{C}}(\vect{k},\vect{r})  
 + \Psi_{\rms{C}}^{S}(\vect{k},\vect{r})
 + \Phi_{\rms{st}}(\vect{k},\vect{r}) 
 + \Phi_{\rms{st}}^{S}(\vect{k},\vect{r})],
\end{equation} 
where superscript $S$ denotes the symmetrization of the wave 
functions.
The functions $ \Psi_{\rms{C}}(\vect{k},\vect{r})$ and  $
\Phi_{\rms{st}}(\vect{k},\vect{r}) $ stand for the Coulomb  wave
function and the wave function induced  by the strong interactions,
respectively.\\

The Coulomb wave function of the identical $\pi$-$\pi$ scattering
with momenta $p_1$ and $p_2$ is given as \cite{schiff,sasagawa}:
\begin{eqnarray}
  \Psi_{\rms{C}}(\vect{k},\vect{r})
  &=& \Gamma(1+i\eta) e^{-\pi\eta/2} 
             e^{i {\vect{\small{k}} \cdot \vect{\small{r}}}}
             F(-i\eta;1;ikr(1-\cos\theta)), \label{Coulomb-wave}
\end{eqnarray}
where $F$ denotes the confluent hypergeometric function, $2k=Q = p_1
- p_2$, ~and \\$\eta = m_{\pi}\alpha/2k $. For calculations of the
confluent hypergeometric function $F$ in Eq.~(\ref{Coulomb-wave}), 
there are two methods. One is to use formulas given 
in Ref.\cite{ratio1}. 
The other is to use the subroutine program `WWHITM ' in the CERN
program library (See Refs.\cite{ratio1,ratio2}).\\

The strong interaction in the Coulomb field is given by
\begin{eqnarray} 
  && \Phi_{\rms{st}}(\vect{k},\vect{r})= \sqrt{G(\eta)}~
     \frac{f_{0}(\theta)\exp[~i(kr-\eta\ln(2kr))~]}{r}, 
     \label{strong-wave}\\ 
  && f_{0}(\theta)=\frac{1}{2ik}
     \exp[~2i~\mbox{arg}\Gamma(1+i\eta)~][~\exp(2i\delta^{(2)}_0)-1~]~. 
     \nonumber 
\end{eqnarray} 
Here $\sqrt{G(\eta)}=\sqrt{\frac{2\pi\eta}{\exp[2\pi\eta]-1}}$ 
is the square root of the Gamow factor which is assumed to be a
normalization factor of the strong wave function\footnote{
This normalization factor is introduced to remove divergence of 
$f_{0}(\theta)$ at $k=0$.} and $\delta^{(2)}_0$ is the the phase 
shift for the $\pi$-$\pi$ ($I$=2) interaction parametrized as
\hspace*{-1mm}\footnote{ 
It is numerically confirmed that calculations for the BEC using 
Eqs. (\ref{Coulomb-wave}) and (\ref{strong-wave}) is equivalent to 
that using a solution of Schr\"odinger equation with Coulomb and 
$\rho$-meson exchange potential\cite{osb96}.}\cite{suzuki87} 
\begin{eqnarray} 
    && \delta^{(2)}_0 = \frac{1}{2}~\frac{a_0^{(2)}Q}{1.0 + 0.5Q^2}~. 
       \label{shift-st}
\end{eqnarray} 
We obtain therefore 
\begin{eqnarray}
      N^{\pm\pm}/N^{BG(mix)} &=& 
      \int_{}^{}\rho(r) d^3r|A_{12}|^2, \nonumber \\
      &=& G(\eta)~[~(1 + \Delta_{\rms{1C}}) + (E_{2\rms{B}} 
      + \Delta_{\rms{EC}}) + I_{\rms{Cst}} + I_{\rms{st}}~], 
\end{eqnarray}
where
\begin{eqnarray} 
   &&  (1 + \Delta_{\rms{1C}}) + (E_{2\rms{B}} + \Delta_{\rms{EC}})
      = \frac{1}{G(k)} 
      \int \!d^3r \rho(r) \Big|\frac{1}{\sqrt{2}}
      [ \Psi_{\rms{C}}(\vect{k},\vect{r})  
      + \Psi_{\rms{C}}^{S}(\vect{k},\vect{r})\Big|^2, \nonumber \\
   &&   E_{2B}=\int \!d^3r \rho(r)
     \exp[-i\mbox{\boldmath Q}\cdot\mbox{\boldmath r}],
     \nonumber \\ 
   &&   I_{\rms{st}}=\frac{1}{G(k)} 
      \int \!d^3r \rho(r) \Big|\frac{1}{\sqrt{2}}
      [ \Phi_{\rms{st}}(\vect{k},\vect{r})  
      + \Phi_{\rms{st}}^{S}(\vect{k},\vect{r})\Big|^2,
\nonumber \\
    &&   I_{\rms{Cst}}=\frac{1}{G(k)} 
       \mbox{Re}\int \!d^3r \rho(r)
        \Big[~\{\Psi_{\rms{C}}(\vect{k},\vect{r})  
      + \Psi_{\rms{C}}^{S}(\vect{k},\vect{r})\}\times
      \{\Phi_{\rms{st}}(\vect{k},\vect{r})  
      + \Phi_{\rms{st}}^{S}(\vect{k},\vect{r})\}^{\ast}~\Big]~, 
\end{eqnarray} 
where $N^{BG(mix)}$ is number of $\pi^+\pi^-$ pairs obtained by event
mixing method. 
In present calculations we use Gaussian source function, 
$\rho(r)=(\frac{1}{\sqrt{2 \pi } \beta})^3 
\exp (\frac{-r^2}{2 \beta^2})$,
the Fourier transform of which is given by:
\begin{eqnarray*}
E_{2 {\rms B}} &=& \exp(-\beta^2 Q^2/2).
\end{eqnarray*}
To obtain the ratio $N^{\pm\pm}/N^{BG(mix)}$ the parameter $\lambda$
should be introduced into it in the usual way. Notice that one more
parameter, the additional normalization factor $c$, is also
introduced by hand. Our final formula with these parameters is then
given as: 
\begin{eqnarray} 
   N^{\pm\pm}/N^{\rms{BG(mix)}} (Q = 2k) 
   \!\!&=&\!\! 
   G(\eta)\times~c( 1 + \Delta_{\rms{1C}} 
   + \Delta_{\rms{EC}}+ I_{\rms{Cst}} + I_{\rms{st}}) \nonumber \\
  & & ~~~~~~~\times 
      \left[ 
           1 + 
     \lambda \frac{ E_{\rms{2B}} }
    { 1 + \Delta_{\rms{1C}}+\Delta_{\rms{EC}}+ 
       I_{\rms{Cst}} + I_{\rms{st}} }
      \right]. 
      \hspace{1cm} \label{eq:ratio}
\end{eqnarray}

It should be noted that the normalization parameter $c$ and an
effective degree of coherence, i.e., the denominator of the ratio
$E_{\rms{2B}}/( 1 + \Delta_{\rms{1C}} + \Delta_{\rms{EC}} + 
I_{\rms{Cst}} + I_{\rms{st}})$, are related to each other.\\

\noindent 
{\bf 3.~~A theoretical formula for $\pi^{+}\pi^{-}$ pairs 
including the FSI:~~}
Using the following amplitudes of the strong interaction for 
the ${\pi^+\pi^-}$ scattering ($I=$2 and 0) and changing 
the sign of the factor $\eta = m\alpha/2q $, 
one can estimate correlations for the ${\pi^+\pi^-}$ pairs. 
The strong amplitude for $\pi^+\pi^- \to \pi^+\pi^-$ scattering 
is decomposed as
\begin{equation}
    \phi_0^{(2,0)}  = \frac{1}{3} f_0^{(2)}(\theta) 
                     + \frac{2}{3}f_0^{(0)}(\theta).\\
\end{equation}

Here, we have assumed that contribution of 
$\pi^0\pi^0 \to \pi^+\pi^-$ channels is small. 
The phase shifts $\delta_0^{(0)}$ for $I = 0$ 
which are obtained in Refs.~\cite{suzuki87}-\cite{phase-6} 
are parameterized by the polynomials (cf., Fig.~1).
\begin{eqnarray}
   && \delta_0^{(0)}(Q)= -75.63~Q + 2345.~Q^2 -12588.~Q^3 
                       +23872.~Q^4 +7431.5~Q^5 \nonumber \\ 
   && \hspace*{2.7cm}    -77590.~Q^6 +87705.~Q^7 -30866.~Q^8~\mbox{[deg]}
                        \quad\mbox{(for $Q\le0.985$ GeV)}\nonumber \\ 
   && \delta_0^{(0)}(Q)=-53061. +179950.~Q -227210.~Q^2 
                        +127110.~Q^3 \nonumber \\ 
   && \hspace*{5.90cm}     -26558.~Q^4 ~\mbox{[deg]} 
           \quad\mbox{(for 0.985$<Q\le1.350$ GeV)}\nonumber \\ 
   &&      \delta_0^{(0)}(Q)=+3978.5 -7988.6~Q +5608.9~Q^2 \nonumber \\
   && \hspace*{5.90cm}    -1269.3~Q^3~\mbox{[deg]}\nonumber 
           \quad\mbox{(for 1.350$<Q\le1.750$ GeV)}\nonumber\\ 
   &&      \delta_0^{(0)}(Q)= 388.05 ~\mbox{[deg]} 
           \hspace*{6.4cm} 
           \quad\mbox{(for 1.750~GeV$ <Q$ )}\nonumber \\ \label{st-pahse}
\end{eqnarray} 
\noindent 
Our formula for the $ {\pi^+\pi^-}$ pairs is given as 
\begin{eqnarray}
      && \Psi_{\rms{total}}(\vect{k},\vect{r}) 
        =\Psi_{\rms{C}}(\vect{k},\vect{r}) +  
         \Phi_0^{(2,0)}(\vect{k},\vect{r}),\nonumber \\
      && \Phi_0^{(2,0)}(\vect{k},\vect{r})=
        ~\sqrt{G(-\eta)}\frac{\phi^{(2,0)}~
         \exp[i(kr+\eta\ln(2kr))]}{r}~,\nonumber \\ 
      && N^{(+-)}/N^{BG(mix)} =
         \int d^3r~\rho(r) ~\vert \Psi_{\rms{total}}(\vect{k},\vect{r})
         \vert^2,\label{plus-minus}
\end{eqnarray} 
It should noted here that Eq.(\ref{plus-minus})  
gives a prediction for $\pi^+\pi^-$ correlation function caused by
the FSI. Using now Eqs. (\ref{eq:ratio}) and (\ref{plus-minus}), one
obtains the following theoretical formula of the BEC with
unlike-particle reference \cite{del97} as
\begin{eqnarray} 
      \pi^{\pm}\pi^{\pm}/\pi^{+}\pi^{-} 
       = \frac{N^{\pm\pm}/N^{BG(mix)}}{N^{+-}/N^{BG(mix)}}~. 
         \label{eq:plus-minus-ratio}
\end{eqnarray} 
~\\

\noindent 
{\bf 4. Analyses of $e^+e^- \to W^+W^-$ events by means of 
Eq.(\ref{eq:plus-minus-ratio}):~~}
We have now analyzed data on $e^+e^- \to W^+W^-$ events at
$\sqrt{s}=172$ GeV \cite{del97} using Eq.(\ref{eq:plus-minus-ratio}). 
A result of minimum-$\chi^2$ fit is shown in Fig.2. Because the
scattering length parameter in Eq.~(\ref{shift-st}) was fixed as 
$a_0=-1.20$GeV${}^{-1}$\cite{suzuki87} and the normalization
parameter $c$ was fixed as $c=1$, we can estimate the remaining two
parameters as:
\begin{center} 
$\bullet$~~
      $\lambda = 1.19 \pm 0.48,~~
      \beta = 0.87 \pm 0.31~\mbox{fm},~~
      (~\chi^2/\mbox{\small NDF} = 14.3/13~)$. 
\end{center} 
It should be noted that the value of the $\lambda$ parameter 
obtained by Eq.(\ref{eq:plus-minus-ratio}) is considerably different 
from one estimated in Ref.\cite{del97,kartvel}. See Table I.\\

In order to expose the cause this difference in a more clear way, 
we also analyze the data by the following two formulas:
\begin{eqnarray}
&&
   \pi^{\pm\pm}/\pi^{+-}(\mbox{strong corr. only}) =~~~
   \frac{N^{\pm\pm}/N^{BG(mix)}}{N^{+-}/N^{BG(mix)}}
   \bigg|_{\mbox{\small $\alpha=0$}}~~, 
   \label{strong-only}\\
&&
   \pi^{\pm\pm}/\pi^{+-}(\mbox{Coulomb corr. only}) =
   \frac{N^{\pm\pm}/N^{BG(mix)}}{N^{+-}/N^{BG(mix)}}
   \bigg|_{\mbox{\small $f_0(\theta)=0$}}~~.
   \label{Coulomb-only}
\end{eqnarray} 
Results of minimum-$\chi^2$ fit to the data by means of Eqs.
(\ref{strong-only}) and (\ref{Coulomb-only}) are summarized in Table 
I$\!$I. We find that the correction caused by the strong FSI  
change drastically the value of the $\lambda$ parameter.
The fitting results by means of Eqs.(\ref{strong-only}) and 
(\ref{Coulomb-only}) are also shown in Fig.3. Since the Coulomb
correction plays important role only in small $Q$ region ($Q\le 0.2
GeV$), strong interaction is dominant in the $Q$ region observed by
authors of Ref.\cite{del97}\footnote{This fact is consistent with
that the Gamow correction for the value of $\lambda$ is small.}.
(cf.~Fig.2 and 3.)\\ 

Finally, we also predict the correlation functions, 
$N^{\pm\pm}/N^{BG(mix)}$ and $N^{+-}/N^{BG(mix)}$ in Fig.4. \\

\noindent 
{\bf 5.~Concluding remarks:~~}
We have analyzed the data of the BEC in ${e^+e^- \to W^+W^-}$ 
events by means of Eq.(\ref{eq:plus-minus-ratio}), i.e., including
full FSI structure (both Coulombic and strong). In this way we have
found that there is normal Bose-Einstein effect in the correlation
data because the degree of coherence parameter $\lambda$ included in
Eq.(\ref{eq:plus-minus-ratio}) comes out to be approximately unity. 
However, this expected enhancements is dissolved by the complicated
structure of the FSI (especially important here due to the small
space-time distances involved in this particular reaction). \\

\noindent 
{\bf Acknowledgments:~~} We are grateful to Dr. A. Tomaradze for
providing us with the experimental data. 
One of us (M. B.) thans the Exchange Program between the JSPS and the
Polish Academy of Sciences and Jagiellonian University, Cracow, for
financial support and acknowledges various conversations with A.
Bialas,  J. Bartke, A. Krzywicki and G. Wilk. This work is partially
supported by  Japanese Grant-in-Aid for Scientific Research from the
Ministry of  Education, Science, Sport and Culture (\# 09440103).
Finally the authors are also indebted to G.Wilk for his reading 
the manuscript. 

\vspace*{4cm}

\noindent 
{\large \bf Figure captions}
\begin{description}

\item
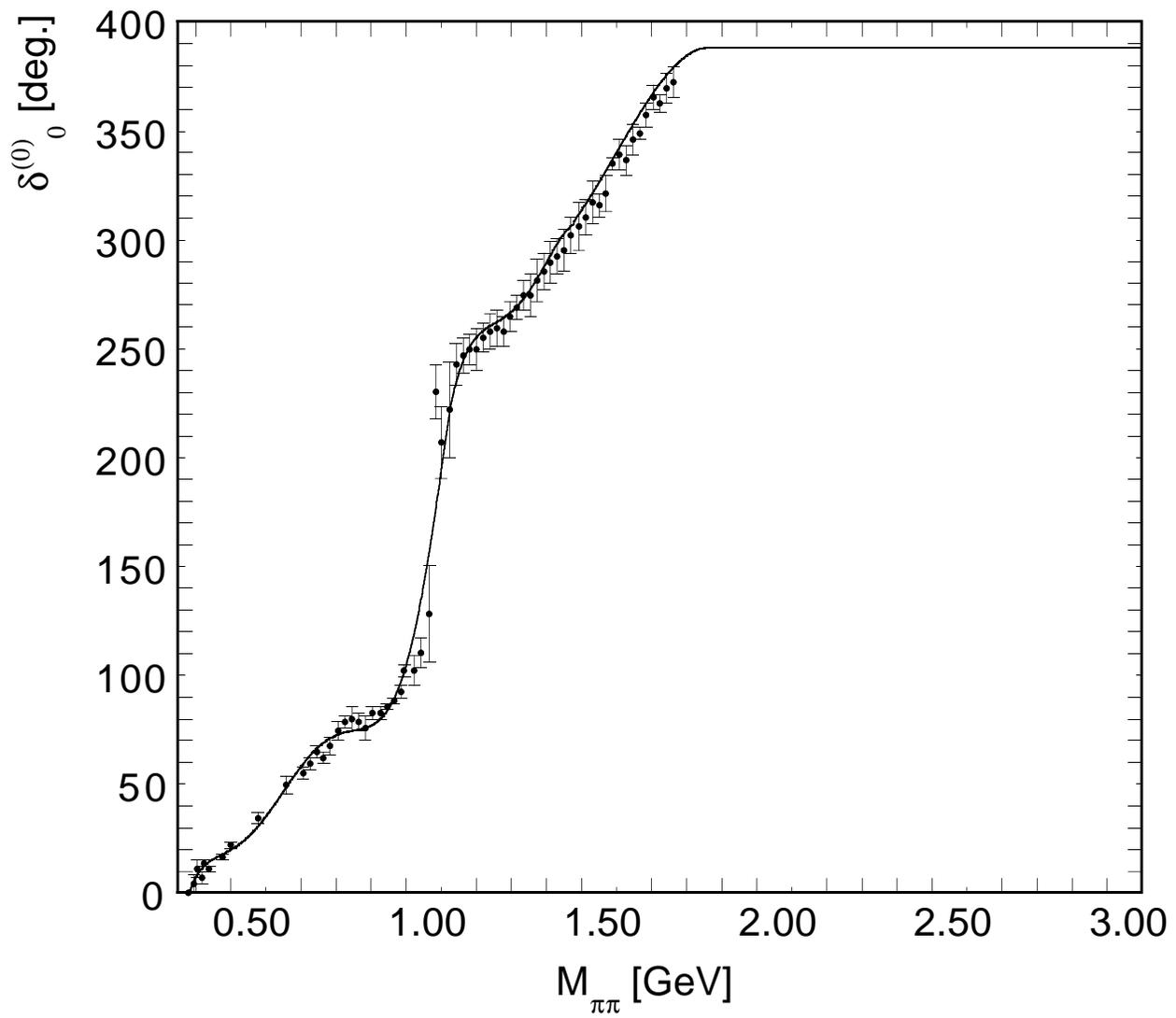
{\bf Fig.1:~} 
The phase shifts of $\delta_0^{(0)}$ parameterized 
by Eq.~(\ref{st-pahse}). See Refs.~[9]-[16].
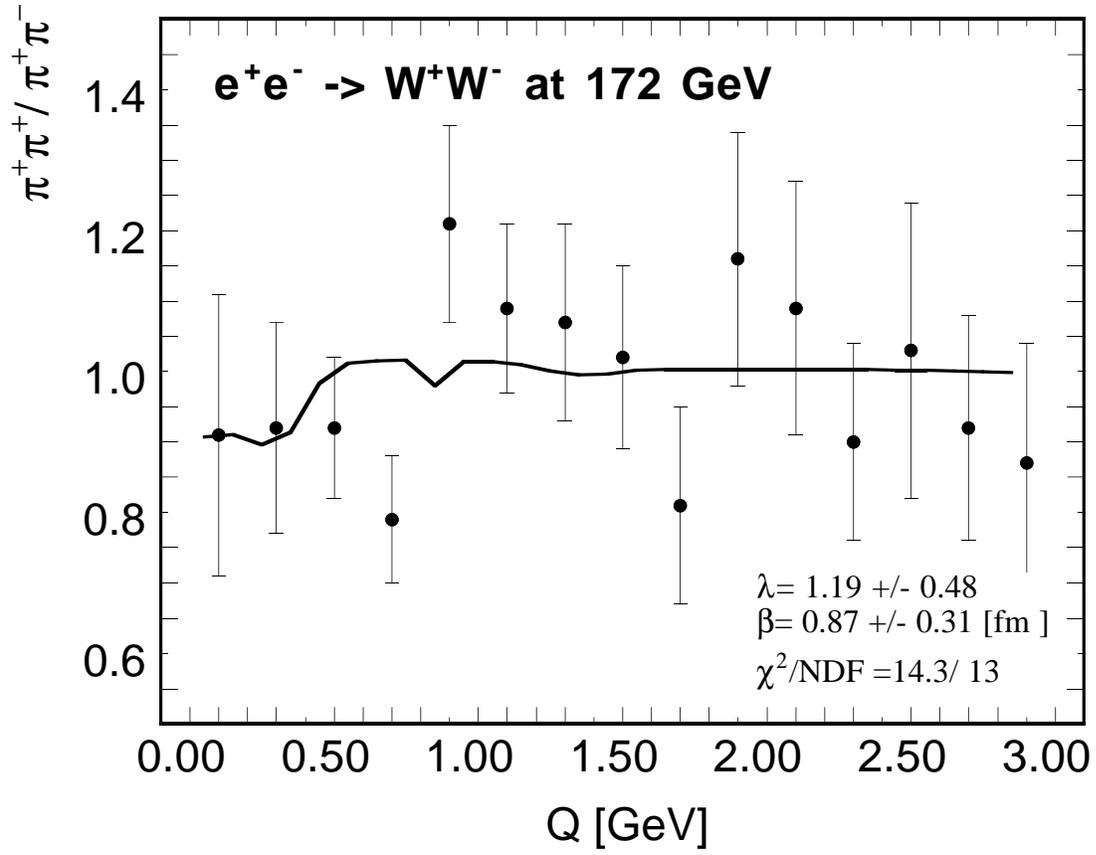
\item {\bf Fig.2:~}
Our result of analysis for the BEC data by 
DELPHI Collab.\cite{del97} using Eq.~(\ref{eq:plus-minus-ratio}). 
\item
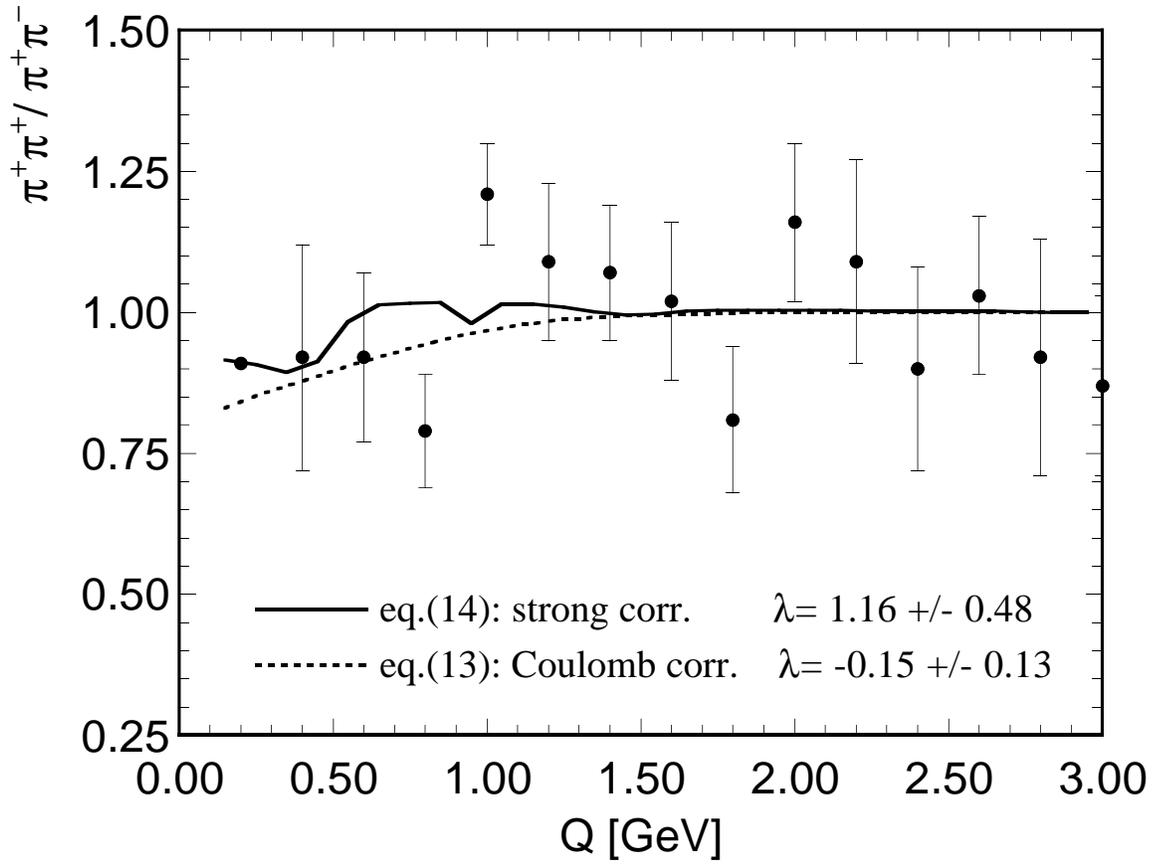
{\bf Fig.3:}
Illustration of the role of the Coulomb and strong final state
interaction in $\pi^{\pm}\pi^{\pm}/\pi^+\pi^-$.
\item
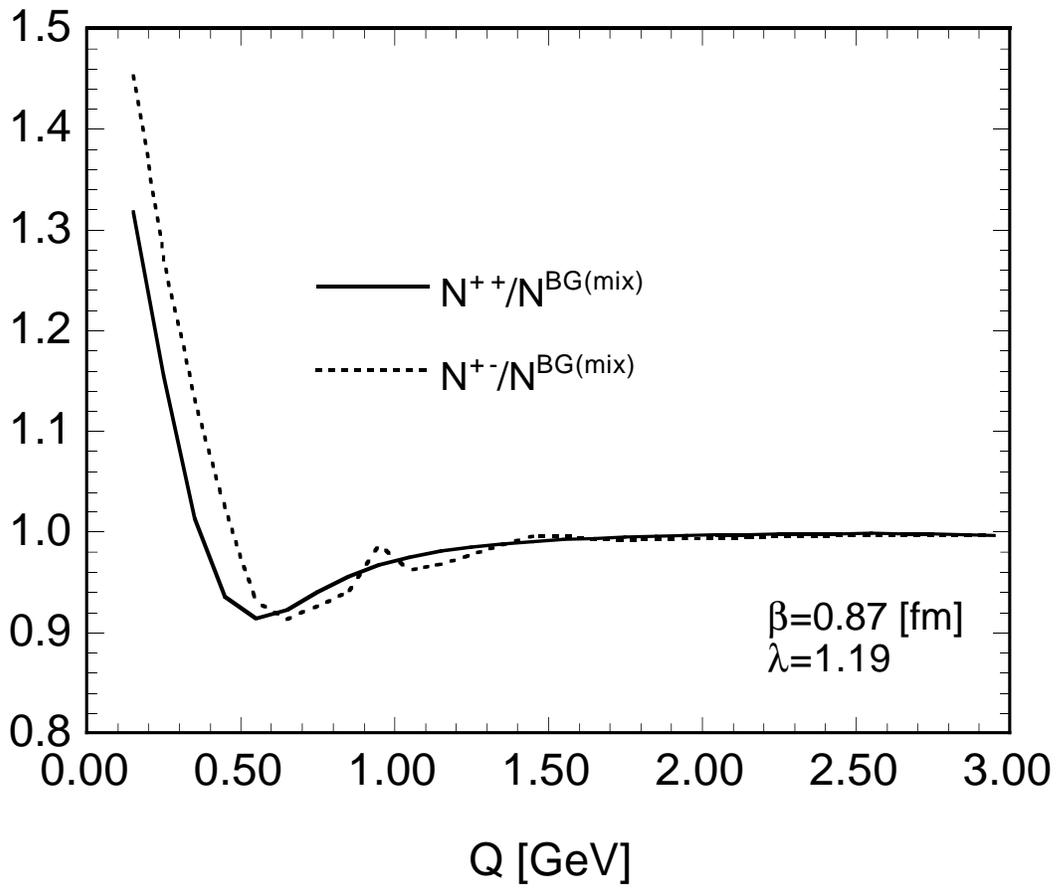
{\bf Fig.4:} Our predictions for correlation of 
$\pi^{\pm}\pi^{\pm}$ pairs (solid) and b) $\pi^{+}\pi^{-}$ pairs (dotted). 
\end{description}

\vspace*{+5mm}
\clearpage 

\noindent 
{\large \bf Table captions}
\begin{description}

\item{\bf Table I:~} Effect of FSI on BEC in $e^+e^- \to W^+W^-$ events. 
\item {\bf Table I$\!$I:~}
Influences of the Coulomb and strong final state interactions 
for the $\beta$ and $\lambda$ parameters.
\end{description}

\vspace*{4cm}
\begin{center} 
\begin{table}[ht]
\vspace*{-0cm}
\begin{tabular}{|l|ll|c|c|}
\hline
Refs.&~$\beta$~[fm]&~$\lambda$& $\chi^2$/NDF& FSI\\
\hline \hline 
Eq.(1) & 0.50~&-0.11 $\pm$ 0.2& $\sim$ 14/14 
& no-FSI\\ 
\lw{Event weighting~[2] }& 
0.322 $\sim$ 0.459 & 0.032 $\sim$ 0.146 &\lw{---}&\\ 
&\multicolumn{2}{c|}{(~depending on weight function~)}&&no-FSI\\ 
Eq.(12) &&&&\\
~~~$\pi^{\pm}\pi^{\pm}/\pi^{+}\pi^{-}$
& 0.86 $\pm$ 0.31 &1.19 $\pm$ 0.48 &14.3/13& Coulomb+ strong\\
\hline 
\end{tabular}
\end{table}
\vspace*{+0mm}
{\large {\bf Table I}}\\
\end{center} 

\vspace*{+10mm}

\begin{table}[ht]
\begin{center} 
\begin{tabular}{|ll|c|c|c|}
\hline
Eq. & FSI & $\beta$~[fm] & $\lambda$ & $\chi^2$/NDF \\ \hline \hline 
Eq.(13) & Coulomb corr. & 0.35 $\pm$ 0.17 & -0.15 $\pm$ 0.13  & 11.5/13 \\
Eq.(14) & strong corr.  & 0.87 $\pm$ 0.32 & 1.16  $\pm$ 0.48  & 14.4/13 \\
\hline
\end{tabular}
\end{center}
\end{table}

\vspace*{-8mm}
\begin{center} 
{\large {\bf Table I$\!$I}}\\
\end{center}
\end{document}